\documentstyle[preprint,epsfig,aps]{revtex}
\begin{document}
%\tighten
%\widetext
\draft
\def\bra#1{{\langle #1{\left| \right.}}}
\def\ket#1{{{\left.\right|} #1\rangle}}
\def\bfgreek#1{ \mbox{\boldmath$#1$}}
\title{In-medium electron-nucleon scattering}
%\footnotemark{}
\footnotetext{ADP97-17-T254}
\author{D. H. Lu, A. W. Thomas, K. Tsushima, A. G. Williams}
\address{Department of Physics and Mathematical Physics\break
 and
        Special Research Centre for the Subatomic Structure of Matter,\break
        University of Adelaide, Australia 5005}
%       Email: dlu@physics.adelaide.edu.au}
\author{K. Saito}
\address{Physics Division, Tohoku College of Pharmacy,\break
Sendai 981, Japan}
\maketitle
\begin{abstract}
In-medium nucleon electromagnetic form factors are calculated 
in the quark meson coupling model. The form factors are typically found to be
suppressed as the density increases. For example, at
normal nuclear density and $Q^2 \sim 0.3 \mbox{ GeV}^2$, 
the nucleon electric form factors are reduced 
by approximately 8\% while the magnetic form factors are reduced 
by only $1 - 2\%$.
These variations are consistent with current experimental limits but
should be tested by more precise experiments in the near future.
\end{abstract}
%\keywords{quark-meson coupling, in-medium nucleon, 
%electromagnetic form factors, y-scaling} 

%\section{Introduction}

There is now considerable evidence that nucleons bound in an atomic
nucleus experience very strong, effective scalar and 
vector fields\cite{QHD,DBHF,Guichon88,Yazaki90,Tony94}. 
It is a fundamental issue in nuclear physics to understand whether these
strong fields alter the internal structure of the nucleon to a
significant extent. There are already strong, experimental constraints
on the extent of such changes\cite{sick,yscaling,coul}. Furthermore, 
new quasi-elastic $(e,e')$ and $(e,e'N)$ experiments at facilities 
such as Mainz, MIT-Bates and especially TJNAF
should improve these constraints in
the near future. Our aim here is to examine the changes in the
electromagnetic form factors of bound nucleons within one particular
quark model of nuclear structure, the quark meson coupling (QMC)
model\cite{Guichon88,Yazaki90,Tony94,Guichon96}. In particular, we show
that the model is consistent with existing constraints but that the
predicted changes should be testable with only a modest increase in
precision.

As in Quantum Hadrodynamics (QHD)\cite{QHD},  
the QMC model describes the properties of nuclear systems using 
effective scalar $(\sigma)$  and vector $(\omega)$ fields.
However, unlike QHD where the nucleon is a point particle, 
the nucleon in this model has substructure.
The $\sigma$ and $\omega$ fields are coupled 
directly to the quarks within the
nucleons, rather than to the nucleons themselves. 
As a result, the internal structure of the bound ``nucleon'' 
is modified by the medium with respect to the free case.
The relative simplicity of the QMC model at the hadronic level, 
together with its phenomenological successes, makes it 
suitable for  many  applications in nuclear physics\cite{BM,JJ,QMC}.
 
The structure of the nucleon in the QMC model is described by 
the MIT bag\cite{MIT}. (However, there are also versions which use the color
dielectric model\cite{NB} or a relativistic oscillator\cite{BM}.)
The small mass of the quark in these models implies that the lower component 
of the quark wave function responds rapidly to the change of 
its environment ($\sigma$ field),
with a consequent decrease in the scalar baryon density. 
As the scalar baryon density itself is the source of the $\sigma$ field, 
this provides a mechanism for the 
saturation of nuclear matter where the quark structure plays a vital role.

In this article we use the self-consistent solutions of the QMC model 
to calculate 
the density-dependent electromagnetic form factors of the nucleon. 
For simplicity, possible off-shell effects of the nucleon 
in medium\cite{Deforest83} are  ignored in the present treatment.
The nucleon bag contains three independent quarks. With the static
spherical cavity approximation to the MIT bag\cite{MIT}, the center-of-mass
motion has to be corrected. We  use the Peierls-Thouless projection method
together with a Lorentz contraction for the nucleon internal 
wave function. This technique for implementing the center-of-mass and
recoil corrections has been quite successful in the case of the 
electromagnetic form factors for free nucleons\cite{Lu97}. 

%\section{Momentum Projection Calculations for A Bare Bag}

For an on-shell nucleon, the electric ($G_E$) and magnetic ($G_M$)
form factors can be conveniently defined in the Breit frame by
\begin{eqnarray}
\bra {N_{s'}({\vec{q}\over 2}) }  J^0(0) \ket {N_s(-{\vec{q}\over 2})} 
&=& \chi_{s'}^\dagger\,\chi_s\, G_E(Q^2), \label{GE} \\
\bra {N_{s'}({\vec{q}\over 2}) } \vec{J}(0) \ket {N_s(-{\vec{q}\over 2})} &=& 
\chi_{s'}^\dagger\, {i\bfgreek\sigma\times\vec{q}\over 2 m_N}\,\chi_s\, 
G_M(Q^2), \label{GM}
\end{eqnarray}
where $\chi_s$ and $\chi^{\dagger}_{s'}$ are Pauli spinors for 
the initial and final nucleons respectively, 
 $\vec{q}$ is the three momentum transfer, 
and $q^2 = -Q^2 = -\vec{q}^{\,2}$. 
The major advantage of the Breit frame is that $G_E$ and $G_M$
are explicitly  decoupled and can be determined by the time and space
components of the electromagnetic current ($J^{\mu}$), respectively.
Note that, in the above definitions [Eqs.~(\ref{GE}) and (\ref{GM}) ], 
both the initial and final states 
are physical states which incorporate meson clouds.

In the QMC model, the quarks typically move much faster than the nucleon 
in the medium (the Fermi motion). Thus it is reasonable to 
assume that they always have 
sufficient time to  adjust their motion so that they stay in the lowest
energy state\cite{Guichon96}.
In the instantaneous rest frame of a nucleon in infinite nuclear matter, 
the Lagrangian for
the quark-meson coupling model is
\begin{eqnarray}
        \protect{\cal L}_q
        &=& \overline q (i\gamma^\mu \partial_\mu - m_q) q\theta_V - B\theta_V
+ g_\sigma^q \overline q q \sigma - g_\omega^q \overline q \gamma_\mu 
q \omega^\mu - {1\over 2} m_{\sigma}^2 \sigma^2 + {1\over 2}
m_{\omega}^2 \omega^2,
\end{eqnarray}
where $m_q$ is the current quark mass, $B$ is the bag constant, 
$g_\sigma^q$ and $g_\omega^q$ denote the quark-meson coupling constants, 
and $\theta_V$ is a step function which is one inside the bag volume
and vanishes  outside.
In mean field approximation for the meson field, 
the normalized solution for the 
lowest state of the quark is given by\cite{MIT} 
\begin{equation}
q_m(t,{\bf r}) =  {N_0\over\sqrt{4\pi}}  e^{-i\epsilon_q t/R} \left (
\begin{array}{c} j_0(\omega_0 r/R) \\ 
i\beta_q \bfgreek {\sigma} \cdot\hat{r} j_1(\omega_0 r/R) \end{array} \right )
\, \theta(R-r)\, \chi_m , \label{cavity}
\end{equation}
where 
\begin{eqnarray}
\epsilon_q &=& \Omega_q + g_\omega\overline{\omega} R, \hspace{1cm}
\beta_q = \sqrt{\Omega_q-m_q^*R \over \Omega_q + m_q^*R}, \\
N_0^{-2} &=& 2R^3 j_0^2(\omega_0)[\Omega_q(\Omega_q-1)+ m_q^*R/2]/\omega_0^2,
\end{eqnarray}
with $\Omega_q \equiv \sqrt{\omega_0^2 + (m_q^*R)^2}$, 
$m_q^* \equiv m_q - g_\sigma^q\overline{\sigma}$, R the bag radius, 
and $\chi_m$ the quark Pauli spinor. The frequency ($\omega_0$)
 of this lowest mode
is determined by the boundary condition at the bag surface,
\begin{equation}
j_0(\omega_0) = \beta_q j_1(\omega_0).
\end{equation}
The form of the quark wave function in Eq.~(\ref{cavity}) 
is almost identical to the free case. 
However the parameters in the solution, Eq.~(\ref{cavity}), 
have been modified by the medium. 
The mean values of the scalar field $\overline{\sigma}$ 
and vector field $\overline{\omega}$ are self-consistently determined by the
coupled equations of motion for the system\cite{Guichon88,Tony94,Guichon96}.
In particular, $\overline{\sigma}$ is given by the 
thermodynamic condition
\begin{equation}
\left.{\partial E_{\rm total} \over \partial \overline{\sigma}}\right|_{R,\rho} = 0,
\end{equation}
where $\rho$ is the baryon density and 
the total energy per nucleon for symmetric nuclear matter is given by
\begin{equation}
E_{\rm total} = {4\over (2\pi)^3\rho}\int^{k_F}\! d^3k\sqrt{{m_N^*}^2+ k^2} 
+ {m_\sigma^2\over 2\rho}\overline{\sigma}^2 
+ {(3 g_\omega^q)^2\over 2m_\omega^2}\rho ,
\end{equation}
and  $\overline{\omega}$ is determined through baryon number conservation to be
\begin{equation}
\overline{\omega} = {3g_\omega^q \over m_\omega^2}\rho\, .
\end{equation}
The effective mass of the nucleon in the medium can be expressed as
\begin{equation}
m_N^* = {3\Omega_q\over R} - { z_0\over R} + {4\over 3}\pi R^3 B,
\end{equation}
where the second term on the right hand side of the above equation parameterizes the sum of the 
center of mass motion
and gluon corrections. 
In the present work, the values of $z_0$ and $B$ are assumed to be 
density-independent (the former for reasons explained in
Ref.\cite{Guichon96}). They are 
determined by requiring that
the free nucleon  mass be given by  $m_N = 939 $ MeV  and by
the equilibrium condition in free space for a given bag radius,
\begin{equation}
\left.{\partial m_N^* \over \partial R}\right|_{\rho=0} = 0. \label{equil}
\end{equation}
For finite nuclear density, the in-medium  bag radius R can be
obtained by solving Eq.~(\ref{equil}) at finite $\rho$ -- it typically
decreases by a few percent at nuclear matter density ($\rho_0$).

The electromagnetic current of the quark is simply
\begin{eqnarray}
j^\mu_q(x) &=& \sum_f Q_f e \overline{q}_f(x) \gamma^\mu q_f(x),\label{current}
\end{eqnarray}
%
%j^{\mu(\pi)}(x) &=& -i e [ \pi^\dagger(x) \partial^\mu \pi(x)
%               -\pi(x) \partial^\mu \pi^\dagger(x)],
%
where $q_f(x)$ is the quark field operator 
for the flavor $f$ and $Q_f$ is its charge in units of e.
The momentum eigenstate of a baryon is constructed 
by the Peierls-Thouless projection method\cite{Lu97,PT62},
\begin{equation}
\Psi_{\rm{PT}}(\vec{x}_1, \vec{x}_2, \vec{x}_3; \vec{p}) = 
N_{\rm{PT}} e^{i\vec{p} \cdot \vec{x}_{\rm{c.m.}} }
q(\vec{x}_1 - \vec{x}_{\rm{c.m.}}) q(\vec{x}_2 - \vec{x}_{\rm{c.m.}})
q(\vec{x}_3 - \vec{x}_{\rm{c.m.}}), \label{PTWF}
\end{equation}
where $N_{\rm{PT}}$ is a normalization  constant, 
$\vec{p}$  the total momentum of the baryon, and 
$\vec{x}_{\rm{c.m.}} = (\vec{x}_1 + \vec{x}_2 + \vec{x}_3)/3$ 
is the center of mass of the baryon (we assume equal mass quarks here).

Using Eqs.~(\ref{current}) and (\ref{PTWF}), the nucleon electromagnetic
form factors for the proton's quark core can be easily calculated by
\begin{eqnarray}
G_E(Q^2) &=& \int\! d^3r j_0(Qr)\rho_q(r)K(r)/D_{\rm PT}, \label{PTE}\\
G_M(Q^2) &=& (2 m_N/Q)\int\! d^3r j_1(Qr) \beta_q j_0(\omega_0 r/R)j_1(\omega_0 r/R)K(r)/D_{\rm PT}, \label{PTM}\\
D_{\rm PT} &=& \int\! d^3r \rho_q(r) K(r),
\end{eqnarray}
where 
$D_{\rm PT}$ is the normalization factor,
$\rho_q(r) \equiv j_0^2(\omega_0 r/R) + \beta_q^2 j_1^2(\omega_0 r/R)$, and 
$K(r) \equiv \int\! d^3x \, \rho_q(\vec{x}) \rho_q(-\vec{x} - \vec{r})$
is the recoil function to account for the correlation of the 
two spectator quarks.

Apart from the center-of-mass correction, it is also vital to include
Lorentz contraction of the bag for the form factors at moderate
momentum transfer\cite{Lu97,LP70}.
In the prefered Breit frame, the photon-quark interaction can be 
reasonably treated as instantaneous. The final form of the form
factors can be obtained through a simple rescaling, i.e.,
\begin{eqnarray}
G_E(Q^2) &=& ({m_N^*\over E^*})^2 G^{\rm sph}_E(Q^2 {m_N^*}^2/{E^*}^2), \\
G_M(Q^2) &=& ({m_N^*\over E^*})^2 G^{\rm sph}_M(Q^2 {m_N^*}^2/{E^*}^2),
\end{eqnarray}
where $E^*=\sqrt{{m_N^*}^2 + Q^2/4}$ and 
$G_{M,E}^{\rm sph}(Q^2)$ are the form factors calculated 
with the static spherical bag wave function [Eqs.~(\ref{PTE}) and (\ref{PTM})].
The scaling factor in the argument arises from the coordinate transformation
of the struck quark and
the factor in the front, $(m_N^*/E^*)^2$,  comes from  the reduction 
of the integral measure of two spectator quarks in the Breit frame\cite{LP70}.

As is well-known, a realistic picture of the nucleon should include
the surrounding meson cloud. 
Following the cloudy bag model (CBM)\cite{CBM,TT83},
we limit our consideration on the meson cloud correction to the 
most important component,  namely the pion cloud.
As in free space, the pion field is a Goldstone boson and acts to restore 
the chiral symmetry.
The Lagrangian related to the pion field and its interaction,
 within the pseudoscalar quark-pion coupling scheme, is
\begin{equation}
\protect{\cal L}_{\pi q} = 
 {1\over 2} (\partial_\mu \bfgreek{\pi})^2
        - {1\over 2} m^2_\pi \bfgreek{\pi}^2
        - {i\over 2f_\pi} \overline q \gamma_5 \bfgreek{\tau} \cdot
        \bfgreek{\pi} q \delta_S, 
\end{equation}
where $\delta_S$ is  a surface delta function of the bag,
$m_\pi$  the pion mass and $f_\pi$  the pion decay constant.
The electromagnetic current of the pion is 
\begin{equation}
j^\mu_\pi(x) = -i e [ \pi^\dagger(x) \partial^\mu \pi(x)
               -\pi(x) \partial^\mu \pi^\dagger(x)],
\end{equation}
where
$ \pi(x) = {1\over \sqrt{2}}[\pi_1(x) + i\pi_2(x)]$
 either destroys a negatively charged pion
or creates a positively charged one.
As long as the bag radius is above 0.7 fm, 
the pion field is relatively weak and can be treated perturbatively. 
A physical baryon state can be expressed as\cite{TT83}
\begin{equation}
\ket A = \sqrt{Z_2^A} [ 1 + (m_A - H_0 - \Lambda H_I \Lambda )^{-1} H_I ] 
\ket {A_0} \label{state},
\end{equation}
where $\Lambda$ is a projection operator which projects out all the components
of $\ket A $ with at least one pion, $H_I$ is the interaction Hamiltonian
which describes the process of emission and absorption of pions.
The matrix elements of $H_I$ between the bare baryon states and 
their properties are\cite{TT83}
\begin{eqnarray}
v^{AB}_{0j}(\vec{k}) &\equiv& 
\bra {A_0} H_I \ket{{\bf \pi}_j(\vec{k}) B_0} = {i f^{AB}_0\over m_\pi}
{u(kR) \over [2\omega_k (2\pi)^3]^{1/2}} \sum_{m,n}
C^{s_B m s_A}_{S_B 1 S_A} (\hat{s}^*_m \cdot {\vec k}) 
C^{t_B n t_A}_{T_B 1 T_A} (\hat{t}^*_n \cdot {\vec e}_j),\\
w^{AB}_{0j}(\vec{k}) &\equiv& 
\bra{A_0 {\bf \pi}_j(\vec{k})} H_I \ket {B_0}
 = \left[v^{BA}_{0j}(\vec{k})\right]^* 
= -v^{AB}_{0j}(\vec{k}) = v^{AB}_{0j}(-\vec{k}),
\end{eqnarray}
where the pion has momentum $\vec{k}$ and  isospin projection $j$,
 $f_0^{AB}$ is the reduced matrix element for the 
$\pi B_0 \rightarrow A_0$ transition vertex, $u(kR) = 3j_1(kR)/kR $,
$\omega_k = \sqrt{k^2+ m^2_\pi}$, and $\hat{s}_m$ and $\hat{t}_n$ 
are spherical unit vectors for spin and isospin, respectively.
The $\pi NN$ form factor [$u(kR)$] is fully determined by the model itself 
and  depends only on the bag radius. 
The bare baryon probability in the physical baryon state, $Z^A_2$, 
is given by
\begin{equation}
Z^A_2 = \left[ 1 +
 \sum_B \left({f^{AB}_0\over m_\pi}\right)^2 {1\over 12\pi^2}
\, \mbox{P}\!\int_0^\infty\! {dk\, k^4 u^2(k R)\over \omega_k (m_A - m_B - \omega_k)^2}
\right]^{-1},
\end{equation}
where P denotes a principal value integral.

Up to one pion loop, the corrections  
to the electromagnetic form factors arising from the pion cloud
are described by the following two processes shown in Fig.~1(b) and 1(c). 
The detailed expressions for their contributions can be 
found in Ref.~\cite{Lu97} with  the following substitutions 
$m_\pi \rightarrow m_\pi^*$, 
$m_B   \rightarrow m_B^*$, 
and $f_{\pi AB} \rightarrow f_{\pi AB}^*$.

In principle, the existence of the $\pi$ and $\Delta$ inside the nuclear 
medium will also lead to some modification of their properties.
% as well as the self-consisitent solution for the nuclear matter. 
Since the pion is well approximated as a Goldstone boson, 
the explicit chiral symmetry
breaking is small in free space, 
and it should be somewhat smaller in nuclear medium\cite{BR91}. 
While the pion mass would be slightly smaller in the medium, 
because the pion field has little
effect on the form factors (other than $G_{\rm En}$), we use $m_\pi^*= m_\pi$. 
As the $\Delta$ is treated on the same footing as the nucleon in the CBM,
its mass should vary in a similar manner as the nucleon. Thus
we assume that the in-medium and free space $N-\Delta$ mass splitting are
approximately equal, i.e., 
$m_\Delta^* - m_N^* \simeq m_\Delta - m_N$.
The physical $\pi AB$ coupling constant is obtained by
$ f^{AB} \simeq \left({f^{AB}_0\over f^{NN}_0}\right) f^{NN}$.
There are uncertain corrections on the bare coupling constant $f^{NN}_0$,
such as the nonzero quark mass and the correction for
spurious center of mass motion.
Therefore, we use the renormalized coupling constant in our
calculation, $f^{NN} \simeq 3.03$, 
which corresponds to the usual $\pi NN$ coupling constant, 
$f^2_{\pi NN}\simeq 0.081$. In the medium, the $\pi NN$ coupling constant
might be expected to decrease  slightly due to the enhancement of the 
lower component of the quark wave function, but we shall
ignore this density dependence in a first treatment
 and use $f_{\pi NN}^* \simeq f_{\pi NN}$.

Fig.~2 summarises the nucleon electromagnetic form factors in free space
with the bag radius R = 1 fm. The `dipole' refers to  the standard dipole fit, 
$F_d(Q^2) = 1/(1 + Q^2/0.71 \mbox{ GeV}^2)^2$.
The corrections for the center-of-mass motion and Lorentz contraction
lead to significantly better agreement with data than was obtained in
the original, static CBM calculations. For a detailed comparison with 
data we refer to Ref.\cite{Lu97}. The purpose of Fig.~2 here is simply to
show that our model produces realistic form factors in free space.

Our main results, namely the  
density dependence of the form factors in matter, 
relative to those in free space, 
are shown in Fig.~3. The  charge form factors are much more sensitive to 
the nuclear medium density than 
the magnetic ones. The latter are nearly one order of magnitude
less sensitive. Increasing density obviously 
suppresses the electromagnetic form factors for small $Q^2$.
For a fixed $Q^2$ (less than $0.3 \mbox{ GeV}^2$), the form factors
decrease almost linearly with respect to the nuclear density, $\rho$\, . 
At $Q^2 \sim 0.3 \mbox{ GeV}^2$, the proton and neutron  charge form factors 
are  reduced by roughly 5\% and 6\% for $\rho = 0.5 \rho_0$, and 
8\% for the normal nuclear density, $\rho_0$; 
similarly, the proton and neutron magnetic form factors are 1\% and 0.6\% 
smaller for $\rho = 0.5 \rho_0$,
and 1.5\% and 0.9\% for the normal nuclear density.

The best experimental constraints on the changes in these form factors
come from the analysis of $y$-scaling data. For example, in Fe the
nucleon root-mean-square radius cannot vary by more than 3\%\cite{sick}. 
However, in
the kinematic range covered by this analysis, the $eN$ cross section is
predominantly magnetic, so this limit applies essentially to $G_{M}$.
(As the electric and magnetic form factors contribute typically in the 
ratio 1:3 the corresponding limit on $G_{E}$ would be 
nearer 10\%.) For the QMC model considered here, the calculated increase 
in the root-mean-square radius of the magnetic form factors is less than
0.8\% at $\rho_0$.
For the electric form factors the best experimental limit seems to 
come from the Coulomb sum-rule, where a variation
bigger than 4\% would be excluded\cite{coul}. 
This is similar in size to the
variations calculated here (e.g., 5.5\% for $G_{\rm Ep}$ at $\rho_0$)
and not sufficient to reject them.

In conclusion, we have calculated the density-dependent electromagnetic
form factors of a bound ``nucleon'' within the QMC model. 
The differential cross section for the  $eN$ scattering 
can be easily constructed once the kinematics is selected. 
This work represents, to our knowledge, 
the first quantitative, density-dependent study 
within a quark model which reproduces the saturation energy, density and
compressibility of nuclear matter. 
While the deviations of the form factors from their free values are
within current experimental limits, they provide a strong motivation
for forthcoming quasielastic electron-nucleus scattering experiments.

We would like to thank I. Sick for helpful correspondence
concerning the constraints from $y$-scaling.
This work was supported by the Australian Research Council. 
A.W.T and K.S. acknowledge 
support from the Japan Society for the Promotion of Science.

\begin{figure}
\vspace{2.5cm}
\centering{\
\epsfig{file=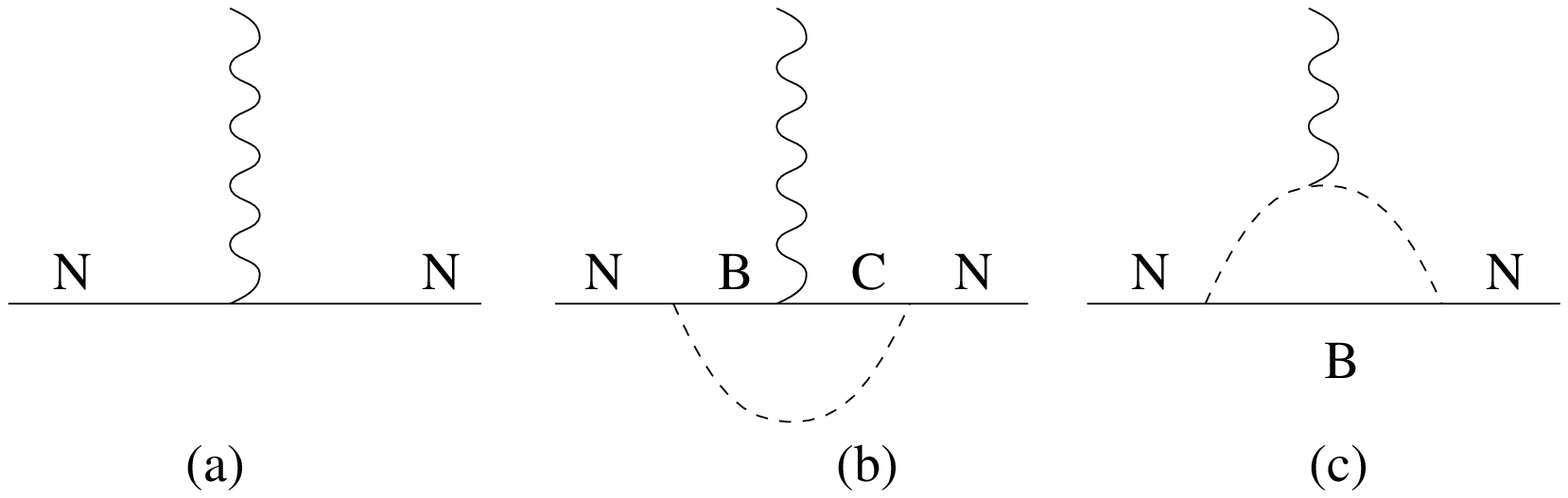,height=4.5cm,width=12cm}
%\vspace*{1cm}
%\epsfbox{fig1.ps}
\vspace{1cm}
\caption{Diagrams illustrating the various contributions included in
this calculation (up to one pion loop). 
The intermediate baryons $B$ and $C$ are restricted
to the $N$ and $\Delta$.}
\label{fig1.ps}}
\end{figure}

\newpage
\vspace*{1cm}
\begin{figure}
\centering{\
\epsfig{file=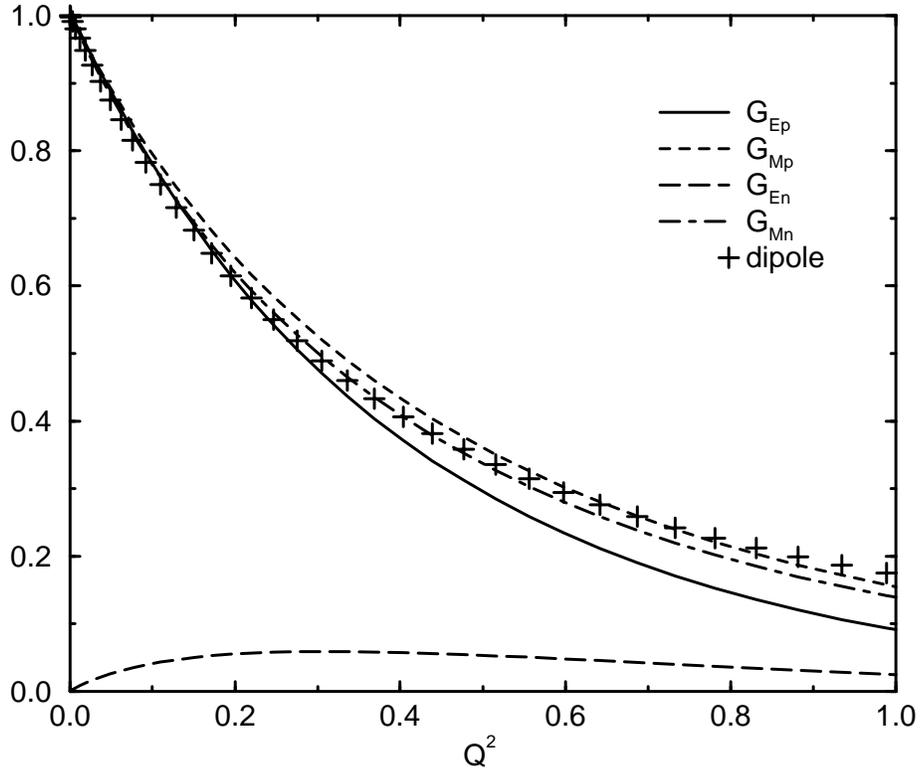,height=12cm}
\caption{The nucleon electromagnetic form factors in free space.
The bag radius was chosen to be R = 1 fm here.}
\label{fig2.ps}}
\end{figure}

\newpage
\vspace*{1cm}
\begin{figure}
\centering{\
\epsfig{file=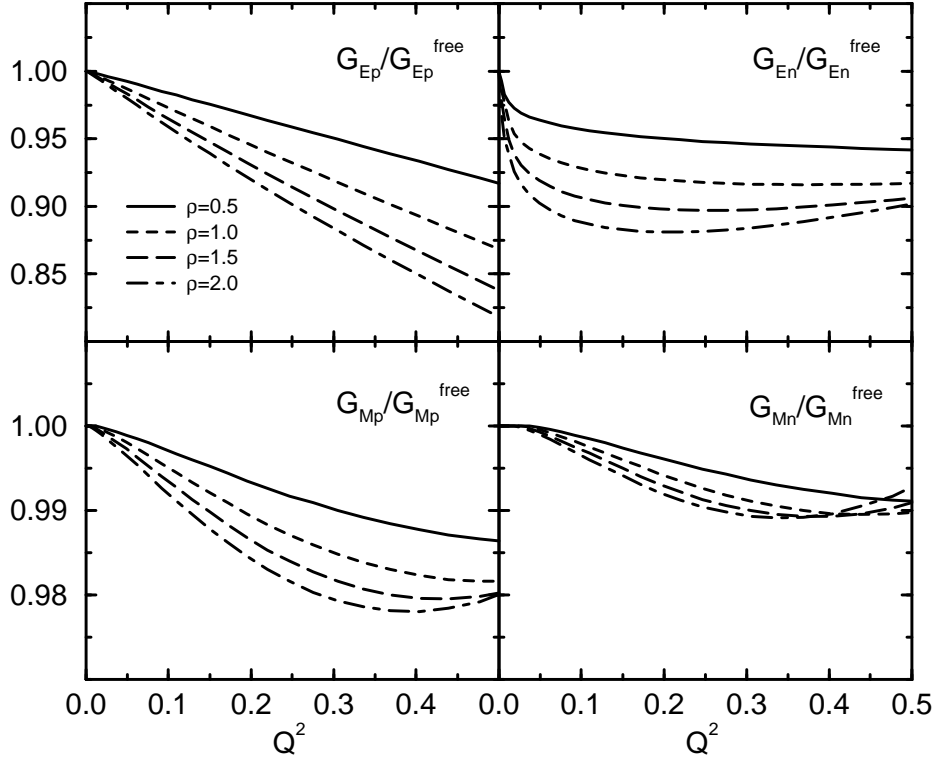,height=12cm}
\caption{The nucleon electromagnetic form factors 
in the nuclear medium (relative to those in free space case).
The free space  bag radius is 1 fm and the density is quoted in units of
the saturation density of symmetric nuclear matter 
($\rho_0 = 0.15 \mbox{ fm}^{-3}$).}
\label{fig3.ps}}
\end{figure}

\end{document}